\documentclass[a4paper,fleqn]{mnras}

\usepackage{mathptmx}
\usepackage{graphics,graphicx}	
\usepackage{amsmath}	
\usepackage{amssymb}	
\usepackage{float}
\usepackage[T1]{fontenc}
\usepackage{ae,aecompl}

\title[A receding torus model for the I-T effect for C-T AGN]
      {A receding torus model for the
  Iwasawa-Taniguchi effect for Compton-thick AGN}

\author[]{
G. Matt$^{1}$,
K. Iwasawa$^{2,3}$
\\
$^{1}$Dipartimento di Matematica e Fisica, Universit\'a degli Studi Roma Tre, via della Vasca Navale 84, 00146 Roma, Italy\\
$^2$Institut de Ci\`encies del Cosmos (ICCUB), Universitat de Barcelona (IEEC-UB), Mart\'i i Franqu\`es, 1, 08028 Barcelona, Spain\\
$^3$ICREA, Pg. Llu\'is Companys 23, 08010 Barcelona, Spain
}

\date{Accepted XXX. Received YYY; in original form ZZZ}

\pubyear{2018}

\begin{document}
\label{firstpage}
\pagerange{\pageref{firstpage}--\pageref{lastpage}}
\maketitle

\begin{abstract}

  Recently, Boorman et al. (2018) reported on the discovery of the
  Iwasawa-Taniguchi (I-T) effect (a.k.a. X-ray Baldwin effect)
  for Compton-thick AGN. They measured a decrease of the
  6.4 keV iron line equivalent width with the 12$\mu$ luminosity,
  assumed as a proxy for the
  intrinsic X-ray luminosity, which in Compton-thick AGN is not directly
  observable. One of the most popular explanations of the classic I-T
  effect is the
  so-called receding torus model, i.e. the decrease of the covering factor of
  the molecular `torus' with X-ray luminosity. In this paper we show that an I-T
  effect for Compton-thick AGN is indeed expected in the receding torus model,
  assuming that the torus is funnelling the primary X-ray luminosity which
  is then scattered in a `hot mirror'. We found that  
  the observed relation is well reproduced provided that the typical column
  density of the `hot mirror' is about 7.5$\times$10$^{22}$ cm$^{-2}$.

\end{abstract}

\begin{keywords}
galaxies: active -- X-rays: galaxies 
\end{keywords}



\section{INTRODUCTION}
\label{intro}

In a recent paper, Boorman et al. (2018) reported on the discovery
of the Iwasawa-Taniguchi (I-T) effect for Compton-thick AGN. The I-T
effect, sometimes also referred to as X-ray Baldwin effect,
is the decreasing of the 6.4 keV iron line equivalent width (EW) with
X-ray luminosity, discovered by Iwasawa \& Taniguchi (1993) and
then confirmed in several other works (e.g. Page et al. 2004; Bianchi et al. 2007;
Ricci et al. 2014). The classic I-T effect, which applies to the narrow
component of the iron line which is likely produced by distant matter (e.g. the pc-scale
'torus'), has been observed in
unobscured sources, where the iron line EW is calculated with
respect to the primary continuum. In Compton-thick AGN, however,
the primary X-ray continuum is completely obscured by intervening
matter, and the I-T effect observed by Boorman et al. (2018) is in
fact against the 12 $\mu$m luminosity, assumed as a proxy for the
intrinsic one. In such sources, the iron line EW
is calculated with respect to the observed X-ray continuum, which
is usually dominated by a reflection component from neutral matter
(here-in-after ``cold reflection''), likely the same matter where
the iron line is originated. {\bf In this case,} 
the line and the reflected continuum should scale in
the same way with the intrinsic luminosity, and the line EW
should therefore be independent of the latter parameter,  if
  properties of the torus like the iron abundance are also independent
  of the intrinsic luminosity, and if the torus is always optically thick.
The I-T effect for Compton-thick source is therefore, at a first
glance, rather surprising.

Among the possible explanations, Boorman et al. (2018) mentioned a
contribution, in the reflection continuum, by a 'mirror' of highly
ionized matter (`hot mirror' here-in-after). In this paper, we explore 
further this hypothesis. Evidence of 
ionized reflection is clearly present in most obscured AGN, expecially
in soft X-rays. This matter may be related to the scattering
mirror which polarizes the broad lines in several Seyfert 2 galaxies
(Antonucci \& Miller 1985; Antonucci 1993; Tran 2001). 
Now, one of the most popular explanations for the classical
I-T effect is the receding torus model, in which the covering factor of the
torus (and therefore the 6.4 keV iron line flux, which is produced {\bf in the
  torus} after reprocessing of the primary emission) decreases with the
luminosity (probably as a result of a dependence on the Eddington ratio, e.g.
Bianchi et al. 2007, Zhuang et al. 2018, even if the latter authors find a
reverse of this dependency for Eddington ratio larger than 0.5).
Let us assume that the ionized gas acting as the hot
reflector fills the opening part of the obscuring torus which naturally
varies as the torus covering fraction changes, and that  
the hot mirror is illuminated by
the intrinsic AGN radiation, which in turn is funneled by the torus.
Under these assumptions, the relative
importance of the hot reflection will increase with luminosity,
providing a possible explanation for the I-T effect for Compton-Thick
AGN. In this way, the latter relation would be a natural consequence
of the classic I-T effect.

In the following, we calculate the expected I-T Compton-Thick effect
in the receding torus model under the above, simple assumptions.
The scope of the paper is just
to check if the proposed explanation could work, and under what conditions,
so for simplicity we will use the observed I-T relations at their face values
without considering the related uncertainties.

\section{The expected I-T effect for Compton-Thick AGN in the receding torus model}

The equivalent width of the iron line in Compton-Thick sources, $EW_{CT}$,
where the primary emission is obscured and the continuum below the line
is entirely due to the reflection components, is given by:

\begin{equation}
  EW_{CT} = {F_{line} \over F_{CR} +  F_{HR} }
\label{eqewct}
\end{equation}

where $F_{line}$ is the iron line flux and $F_{CR}$ and $F_{HR}$ are
the continua at 6.4 keV due to cold (torus) and hot (mirror)
reflection, respectively.

On the other hand, Bianchi et al. (2007) established the I-T relation for
unobscured AGN:

\begin{equation}
EW = {F_{line} \over F_{int} } = A \alpha(L_x) 
  \end{equation}
    
where $F_{int}$ is the continuum flux at 6.4 keV due to the intrinsic
radiation (which we assume to be, at that energy, dominant with respect
to both reflection continua). $\alpha$ represents the luminosity-dependent
fractional covering factor of the torus. $A$, therefore, is the EW for a 
covering factor of order unity, i.e. $\sim$110 eV (e.g. Ghisellini, Haardt
\& Matt; Matt, Guainazzi \& Maiolino 2003). From
Bianchi et al. (2007), $\alpha$ can be written as 0.5$L_{x, 44}^{-0.17}$,
$L_{X, 44}$ being the 2-10 keV luminosity normalized to 10$^{44}$ erg/s. 
This implies that the relation holds only down to $L_{X, 44} \sim$0.01;
below this value, it has to saturate,  otherwise the covering factor would
  exceed unity.).

The cold reflection component, being due to the same matter producing the
iron line, should follow the same luminosity dependence,
$F_{CR} = B \alpha(L_X)F_{int} $, where $B$ is the ratio
between reflected and intrinsic fluxes for a covering factor of order unity.
In our model, the hot mirror is illuminated by the radiation funnelled by the
torus, and therefore its covering factor is complementary to that of the torus 
itself. Therefore, the hot reflection component is 
given by $F_{HR} = C [1 - \alpha(L)]F_{int}$, where $C$ again is the ratio
between reflected and intrinsic fluxes for a covering factor of order unity.
 Inserting in Eq.~\ref{eqewct}, and after simple algebra:

\begin{equation}
  EW_{CT} = { 1 \over { {B \over A} +{C \over A \alpha(L)} - {C \over A} } } 
\end{equation}

\begin{figure}
\includegraphics[width=1.0\columnwidth]{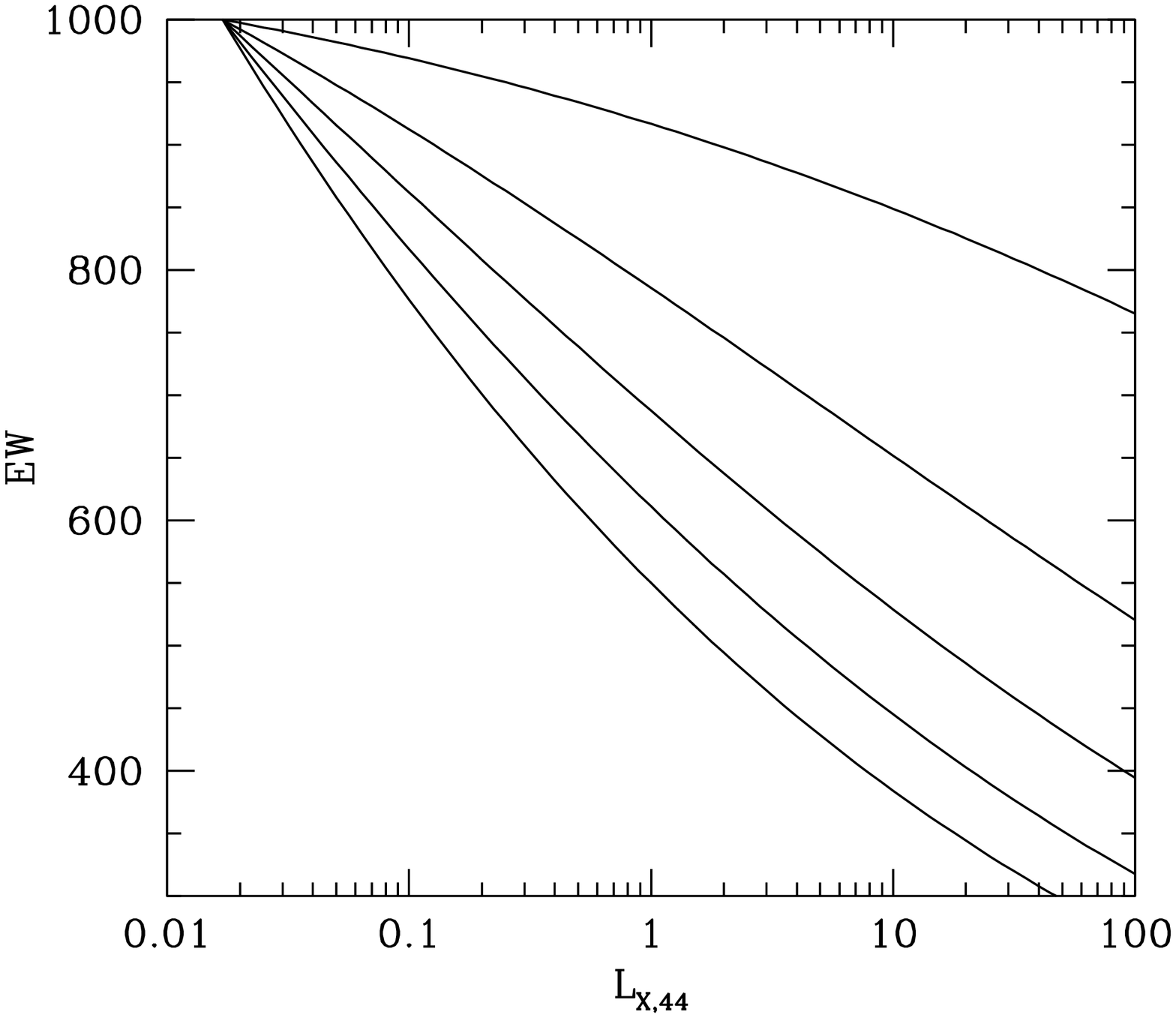}
\includegraphics[width=1.0\columnwidth]{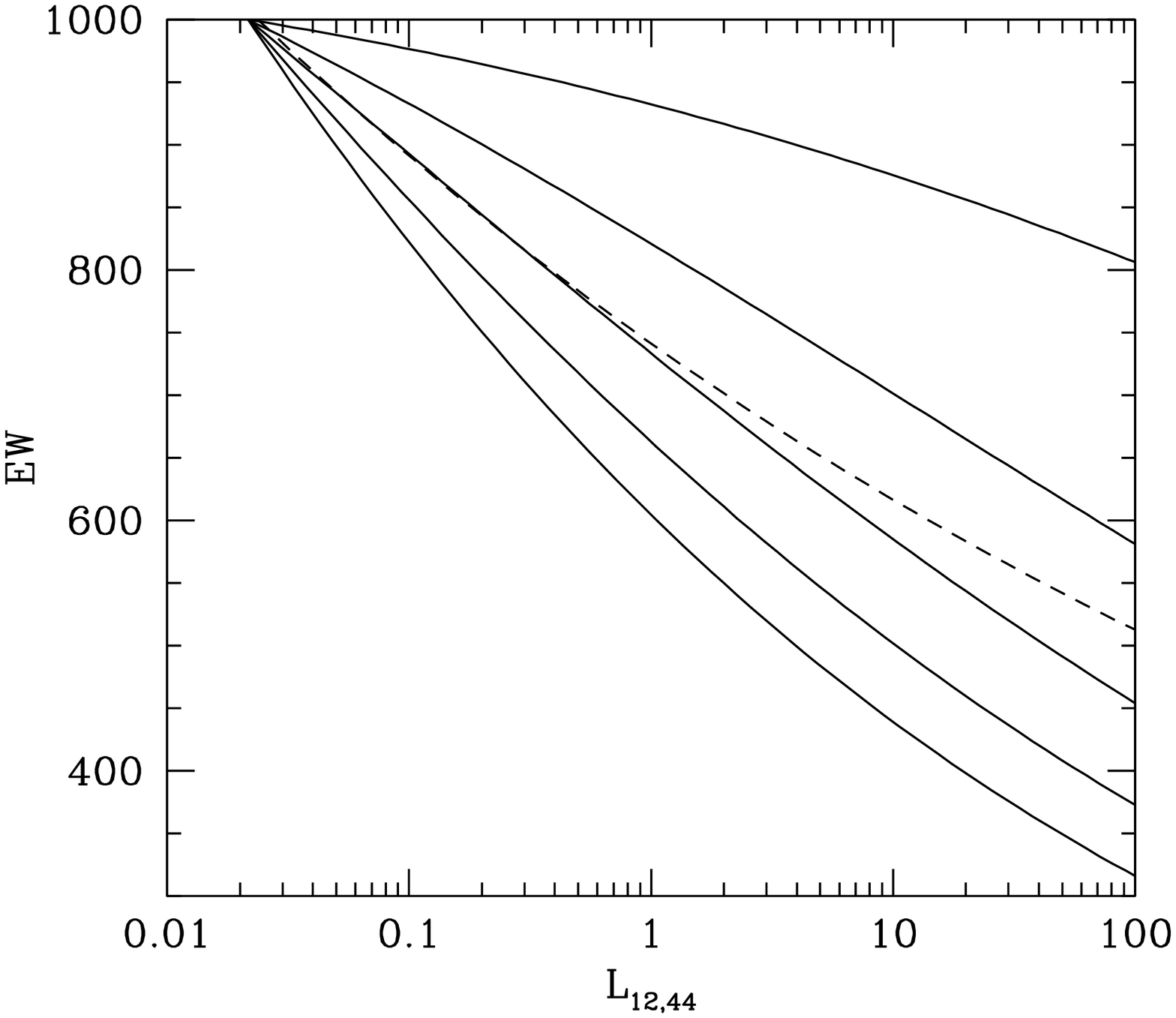}
\caption{The iron line equivalent width for Compton-thick
  sources as a function of the 2-10 keV luminosity (upper panel)
  and of the 12$\mu$m luminosity (lower panel). The five curves
  refer to five values of $C$=0.01, 0.03, 0.05, 0.07, 0.09 (from top to bottom).
  The dashed line in the lower panel is the best-fit relation of
  Boorman et al. 2018. See text for details.
  }
 \label{ITCT}
\end{figure}

Here, $B/A$ corresponds to $F_{line}/F_{CR}$ and is essentialy the iron line equivalent width
with respect to Compton
reflection only, which we assume to be $\sim$1 keV  (e.g. Ghisellini, Haardt
\& Matt; Matt, Guainazzi \& Maiolino 2003). At this point, the I-T effect
for Compton-thick sources can be written as a function of the intrinsic luminosity,
assuming a certain value for $C$, which is the fraction of scattered radiation
from the hot mirror for a covering factor of 1, and thence represents the
Thomson optical depth of the reflecting matter 
($C$=1 would then correspond to a column density, $N_H$, of $\sim$1.5$\times$10$^{24}$ cm $^{-2}$,
but note that the relation strictly holds only for optically thin matter). 
The expected $EW_{CT}$ vs. intrinsic luminosity is
shown in Fig.~\ref{ITCT} (upper panel) for different choices of $C$.

The I-T effect found by Boorman et al. (2018) is, however, as a function of the 12$\mu$ luminosity.
In the lower panel of  Fig.~\ref{ITCT}, the  $EW_{CT}$ versus this luminosity is plotted,
adopting the relation between nuclear (e.g. AGN-dominated) 
 mid-infrared and X-ray luminosities reported by Gandhi et al. (2009) in their eq. 2. 

\section{Discussion and conclusions}

From Fig.~\ref{ITCT}, lower panel, it can be seen that a good agreement with the
observed relation is found for $C$=0.05  ($N_H$=7.5$\times$10$^{22}$ cm $^{-2}$), 
even if the theoretical and observed curves
tend to diverge at high luminosities, where however there are not many data points. 
Because $C$ is {\it the scattering fraction for a covering factor of
  1 of the hot mirror}, this implies that the luminosity-dependent scattering fraction
(which is the product of the opacity, which we assume to be the same for all sources, and
the luminosity-dependent covering fraction)
goes from 5\% at very large luminosities down to fractions of percent for luminosities
around 10$^{42}$ erg/s, where the relation has to saturate, as mentioned in the previous
section. 

Not much is known about the scattering fraction of the hot mirror, observationally, due also
to the difficulties, in Compton-thick sources at least, to assess the intrinsic luminosity.
Moreover, contamination from thermal emission may be
significant for low lumimosity sources and/or in the presence of intense star-forming regions
(e.g. Wang et al. 2014).
A systematic study of the luminosity dependence of the hot mirror scattering efficiency would be
required to test the model presented in this paper. At present, information is rather sparse.
High values (3-6\%) have been found in NGC~7674 (Gandhi et al. 2017), a source with an intrinsic
2-10 keV X-ray luminosity likely in the 0.1--1$\times$10$^{44}$ erg/s range. In NGC~1068, on the other
hand, the value seems to be well below 1\% (Matt et al. 2004), and the same is true for several other sources
(e.g. Ueda et al. 2007; Comastri et al 2010; Eguchi et al. 2011). Ricci et al. (2017) measured this
parameter in a handful of Swift-BAT selected sources finding values ranging from $<$0.7 to 2.  

A potential problem is that ionized iron lines may also be produced in the hot mirror, increasing
the 6.4 keV iron line equivalent width for the faintest sources for which 
a clear separation between neutral (6.4 keV) and ionized
(6.7-7 keV) lines may be difficult even with CCD-like detectors. On the other hand, 
the presence of such lines may help explaining the flattening of the observed curve at high luminosities
(and therefore, in our model, at high scattering fractions) with respect to the calculated one.
He- and H-like iron lines have indeed been observed in some (but not all) Compton-thick AGN, most notably NGC~1068
(Iwasawa, Fabian \& Matt 1997; Guainazzi et al. 1999; Matt et al. 2004; Bauer et al. 2015). 
Matt, Brandt \& Fabian
(1996) calculated the iron line equivalent width expected in hot mirrors. For the column densities
required to explain the I-T effect for Compton-thick AGN, i.e. well above 10$^{22}$ cm$^{-2}$,
a significant suppression of the
line flux is expected due to resonant scattering. Moreover, the hot mirror may be either
mildly or completely ionized, and in both cases no significant line emission
is expected (in the
mildly ionized case, again due to resonant trapping, Ross \& Fabian 1993). It is also
worth mentioning that the cold reflection component in Compton-thick sources is likely
to pass through the hot mirror, and high ionization iron absorption lines may be
imprinted in the reflection spectrum. With present, low
resolution spectrometers it is hard to assess the physical conditions of the hot mirror,
expecially for relatively faint sources. A breakthrough in this respect is expected
with high spectral resolution detectors, like those on-board XRISM and Athena.

To summarize, we have shown that the receding torus model, often invoked to explain the classic
I-T effect, may also naturally explain the recently discovered I-T effect for Compton-thick AGN
(Boorman et al. 2018),
provided that the hot mirror has typically a column density of $\sim$7.5$\times$10$^{22}$ cm$^{-2}$.
High quality, high energy resolution observations of Compton-thick sources are required
to check the validity of our assumptions.

\section*{Acknowledgements}

We thank Stefano Bianchi for useful discussions and the referee, 
Franz Bauer, for valuable comments which helped improving the clarity
of the paper. 
This work has been started during a visit, under the Maria de Metzeu program,
of 
GM to the University of Barcelona, whose hospitality he gratefully acknowledges.
KI acknowledges support by the Spanish
MINECO under grant AYA2016-76012-C3-1-P and MDM-2014-0369 of ICCUB
(Unidad de Excelencia 'Mar\'ia de Maeztu').










\bsp	
\label{lastpage}
\end{document}